\documentclass[12 pt, amsfonts,amsmath, amssymb,color]{article}

\usepackage{amsmath,amssymb,amsfonts,latexsym,float,graphics,epsfig}
\usepackage{subfig}
\usepackage{verbatim}
\usepackage{relsize}

\def\be{\begin{equation}}
\def\ee{\end{equation}}
\def\bea{\begin{eqnarray}}
\def\eea{\end{eqnarray}}

\begin{document}
%<<<<<<<<<<< enumeration of eqns section wise>>>>>>>>>>>>>>>>>>>

\renewcommand\theequation{\arabic{section}.\arabic{equation}}
\catcode`@=11 \@addtoreset{equation}{section}
%<<<<<<<<<<<<<<<<<<<<<<<<<<<<<<<<<>>>>>>>>>>>>>>>>>>>>>>>>>>>>>>>>>
\newtheorem{axiom}{Definition}[section]
\newtheorem{theorem}{Theorem}[section]
\newtheorem{axiom2}{Example}[section]
\newtheorem{lem}{Lemma}[section]
\newtheorem{prop}{Proposition}[section]
\newtheorem{cor}{Corollary}[section]

\newcommand{\ben}{\begin{equation*}}
\newcommand{\een}{\end{equation*}}
\title{\bf Action-angle variables for the purely nonlinear oscillator}
\author{
\bf Aritra Ghosh\footnote{E-mail: ag34@iitbbs.ac.in} \hspace{0.5mm} and Chandrasekhar Bhamidipati\footnote{E-mail: chandrasekhar@iitbbs.ac.in}\\
~~~~~\\
 School of Basic Sciences, Indian Institute of Technology Bhubaneswar,\\   Jatni, Khurda, Odisha, 752050, India\\
}

\date{ }

\maketitle

\begin{abstract}
In this letter, we study the purely nonlinear oscillator by the method of action-angle variables of Hamiltonian systems. The frequency of the non-isochronous system is obtained, which agrees well with the previously known result. Exact analytic solutions of the system involving generalized trigonometric functions are presented. We also present arguments to show the adiabatic invariance of the action variable for a time-dependent purely nonlinear oscillator.
\end{abstract}

\smallskip


% \maketitle
\section{Introduction}
In recent times, there has been a considerable amount of interest in strongly nonlinear oscillators [1-8]. In particular, oscillators with a restoring force proportional to \(sgn(x)|x|^\alpha\) with \(\alpha\) being a positive rational number have attracted much attention lately. These dynamical systems are interesting both from a theoretical and an applications perspective. Such oscillators arise in various applications in engineering and mechanics where the nonlinearity associated with the system need not be associated with an integer power in the restoring force.

From an applications perspective ([6] and references therein), nonlinear oscillations occur in several physical systems of various length scales ranging from macroscopic scales to nano scales. These are ubiquitous in nature and arise not only in mechanics by also in electronics, biological systems etc. Experimental investigations on aircraft materials, various alloys, wood, ceramic materials, hydrophilic polymers, composites etc reveal that in many real life situations the stress-strain properties of the material are strongly nonlinear i.e., the variation of force with displacement or deflection is quite rapid. In such a situation, the conventional polynomial approximation of the nonlinear restoring force is often not quite useful and this indicates the importance of the generalized model of the purely nonlinear oscillator where the restoring force is not necessarily an integer power of the displacement.

When the restoring force was an odd integer power of the displacement, one can in principle express solutions in terms of Jacobi elliptic functions (see for example [9,10]). The inclusion of the signum function in this model allows for much general nonlinearity with even integer and other rational powers in the restoring force. For this model however, solutions can be expressed in terms of ateb functions [11-13] which are inversions of the incomplete beta functions defined as follows,
\begin{equation} \label{betaa}
B_x(a,b) = \int_{0}^{0\leq x \leq 1}(x')^{a-1}(1-x')^{b-1} dx' \, .
\end{equation}
Senik [12,13] showed that the ateb functions are the solutions of the coupled system,
$$ \dot{x} = y^{\alpha}, \qquad \dot{y} = - \frac{2}{\alpha + 1} x, $$
namely $x(t) = sa(1,\alpha,t)$ and $y(t) = ca(\alpha,1,t)$. They satisfy the identity: $$sa^2(\alpha, 1, t)+ca^{\alpha+1}(1, \alpha, t)=1$$ and are \(2\Pi_\alpha\) periodic with \begin{equation}\Pi_\alpha:=B\bigg(\frac{1}{\alpha+1}, \frac{1}{2}\bigg) = \frac{\Gamma(\frac{1}{2})\Gamma\left(\frac{1}{\alpha+1}\right)}{\Gamma\left(\frac{\alpha+3}{2(\alpha+1)}\right)}\end{equation}. They clearly resemble the circular sine and cosine functions and are known as ateb sine \(sa\) and cosine \(ca\) functions in the literature (see [11-13] for details; Appendix A of [6]). For the case \(\alpha=1\) these reduce to the regular sine and cosine functions.

The solution of the purely nonlinear oscillator, $$\ddot{x}+sgn(x)|x|^\alpha=0$$ subject to the initial conditions $x(0)=A$ and $\dot{x}(0)=0$ is written as [6],$$x(t)= A ca(\alpha, 1, \omega_{ca}t),$$ with the frequency \(\omega_{ca}\) given by \begin{equation}\omega_{ca}=\sqrt{\frac{\alpha+1}{2}}|A|^{(\alpha-1)/2}\end{equation}. The period function is given as [6,7]: \begin{equation}T(A)=\sqrt{\frac{8\pi}{\alpha+1}}
\frac{\Gamma\left(\frac{1}{\alpha+1}\right)}{\Gamma\left(\frac{\alpha+3}{2(\alpha+1)}\right)}|A|^{(1-\alpha)/2}\end{equation}.

It can be verified that, \begin{equation}\frac{2\Pi_\alpha}{\omega_{ca}} = T(A)\end{equation}.

The frequency (and period function) depend on the amplitude \(A\) of the system and hence exhibit non-isochronous behavior. All the results for this system reduce to those of the harmonic oscillator with \(\alpha=1\) wherein, the frequency becomes independent of amplitude and solutions become the standard trigonometric functions.\\

\textbf{Motivation and results:}
In this paper, we use an entirely different approach to solve the purely nonlinear oscillator, which are known as action-angle variables. Action-angle variables are well known in literature for periodic systems and they have also thrown enormous insights in to quasi-periodic systems where the corresponding Hamiltonian system is time-dependent.
In situations, where the frequency or length scale of the problem is a time-dependent function and the system is quasi-periodic, it is known that action-angle variables provide the best approach to obtain adiabatic invariants of the system, which remain unchanged during the motion.  It is clear that this system is a generalization of the linear harmonic oscillator to a nonlinear system with restoring force being an arbitrary positive power of the displacement. For \(\alpha=1\), the system reduces to the harmonic oscillator and hence, the solutions are then the standard (circular) trigonometric functions. The solutions we obtain are in terms of generalized trigonometric functions being defined [14-18] as parametric equations to the curve,

\begin{equation}
  |x|^m + |y|^n = 1,  \hspace{10mm}      m,n \in (1,\infty)
\end{equation}
with, \(y= \sin_{m,n} \theta\) and \(x = \cos_{m,n} \theta\) which are the generalized sine and cosine functions respectively corresponding to parameters \(m\) and \(n\). Here, \(m\) and \(n\) are two arbitrary rational number parameters greater than one (see Appendix B for a more detailed discussion on these functions). However for the physical model of the purely nonlinear oscillator, the solutions are in terms of generalized trigonometric functions with \(m=2\) and \(n=\alpha+1\) (see below).

\smallskip

The main result of this paper may be stated as follows,

\begin{theorem}
For the purely nonlinear oscillator given by the equation of motion \begin{equation}\ddot{x} + sgn(x)|x|^\alpha = 0\end{equation} subject to the initial conditions $x(0) = A$ and $\dot{x}(0) = 0$, the solution may be expressed as \begin{equation}x(t) = A \sin_{2,\alpha+1} (\Omega_{2,\alpha+1}t+\pi_{2,\alpha+1}/2)\end{equation} with \begin{equation}
  \Omega_{2,\alpha+1} = \sqrt{\frac{2}{\alpha+1}}|A|^\frac{\alpha-1}{2}
\end{equation}
\end{theorem}

The structure of the paper is as follows. In the following section, we define the action-angle variables for the system. We reproduce the previously known [6,7] result of the period function of the system by the action-angle method. Exact solutions of the system are then presented in terms of generalized trigonometric functions. In section 3, we relate the periodicity of these the two classes of functions and remark on connections between the two special functions. We end with remarks showing the adiabatic invariance of the action variable for purely nonlinear oscillator. Brief appendices on action-angle variables and generalized trigonometric functions are also included.

\section{The purely nonlinear oscillator}
We consider the dimensionless purely nonlinear oscillator described by the equation of motion,

\begin{equation}
  \ddot{q} + sgn(q)|q|^{\alpha} = 0
\end{equation}

or the allied system,
  $$\ddot{q} + q|q|^{\alpha-1} = 0$$

subject to initial conditions, \(q(0)=q_0\) and \(\dot{q}(0)=0\).

\smallskip

The signum function ensures that the restoring force is an odd function of the displacement. The equation of motion can be derived from the potential,
$$V(q) =  \int |q|^{\alpha}dq = \frac{|q|^{\alpha + 1}}{\alpha + 1}$$

The system admits a Hamiltonian structure with the Hamiltonian function taking the form,

\begin{equation} \label{HNA}
  H(q,p) = \frac{p^2}{2} + \frac{|q|^{\alpha + 1}}{\alpha + 1}
\end{equation}

This is equal to the total energy of the system, \(H(q,p) = E\) which is a constant of motion.

\subsection{The action-angle variables}
Here we consider defining the action-angle variables (see Appendix A). Solving \(H(q,p) = E\) for \(p = p(q)\),
\begin{eqnarray}
 \nonumber
  p &=& \sqrt{2E} \bigg(1 - \frac{|q|^{\alpha+1}}{(\alpha + 1)E}\bigg)^{1/2} \\
  \nonumber
   &=& \sqrt{2E} \bigg(1 - {\bigg|\frac{q}{q_0}\bigg|^{\alpha+1}}\bigg)^{1/2}
\end{eqnarray}

The action can be defined as:

\begin{equation}\label{2.1.1}
  J = \oint pdq
\end{equation}

Therefore one gets

\begin{eqnarray}
 \nonumber %Remove numbering (before each equation)
  J &=& \sqrt{2E} \oint \bigg(1 - {\bigg|\frac{q}{q_0}\bigg|^{\alpha+1}}\bigg)^{1/2}dq \\
  \nonumber
   &=& \sqrt{2E} q_0 I_{\alpha+1}
\end{eqnarray}

\smallskip

where, \(I_{\alpha+1}\) is an integral independent of \(q_0\) defined as,

\begin{equation}\label{2.1.2}
  I_{\alpha+1} = \oint \sqrt{1 - |\xi|^{\alpha+1}}d\xi
\end{equation}

with \(\xi=\frac{q}{q_0}\).

\vspace{2mm}

We now introduce angular coordinate \(\theta\) as,

\begin{equation}\label{2.1.3}
  \theta = \int_{0}^{\xi} \frac{dx'}{\sqrt{1 - |x'|^{\alpha+1}}} := \arcsin_{2,\alpha+1} \xi
\end{equation}

where, \(\arcsin_{m,n} (.)\) is an inversion of the generalized sine function [14].\\

Using the identity, $|\sin_{m,n} \phi|^n + |\cos_{m,n} \phi|^m = 1$ and that $\cos_{m,n} \phi := \frac{d}{d\phi} (\sin_{m,n} \phi)$, one expresses the integral as:

\begin{equation}\label{2.1.4}
  I_{\alpha+1} = \int_{0}^{2\pi_{2,\alpha+1}}  \cos^2_{2,\alpha+1} \theta d\theta
\end{equation}\\

where, \(\pi_{2,\alpha+1}\) is a generalized pi defined as:

\begin{equation}
\nonumber
  \pi_{m,n} := 2\int_{0}^{1} \frac{dx'}{\sqrt[n]{1 - |x'|^m}}
\end{equation}\\

This integral [eqn (2.6)] can be expressed as:

\begin{equation}\label{2.1.5}
I_{\alpha+1} = \frac{4}{\alpha+3}B\bigg(\frac{1}{2},\frac{1}{\alpha+1}\bigg)
\end{equation}\\
\begin{figure}[h]
%	 \begin{wrapfigure}{l}{0.3\textwidth}
	\begin{center}
		\centering
		\includegraphics[width=4.2in]{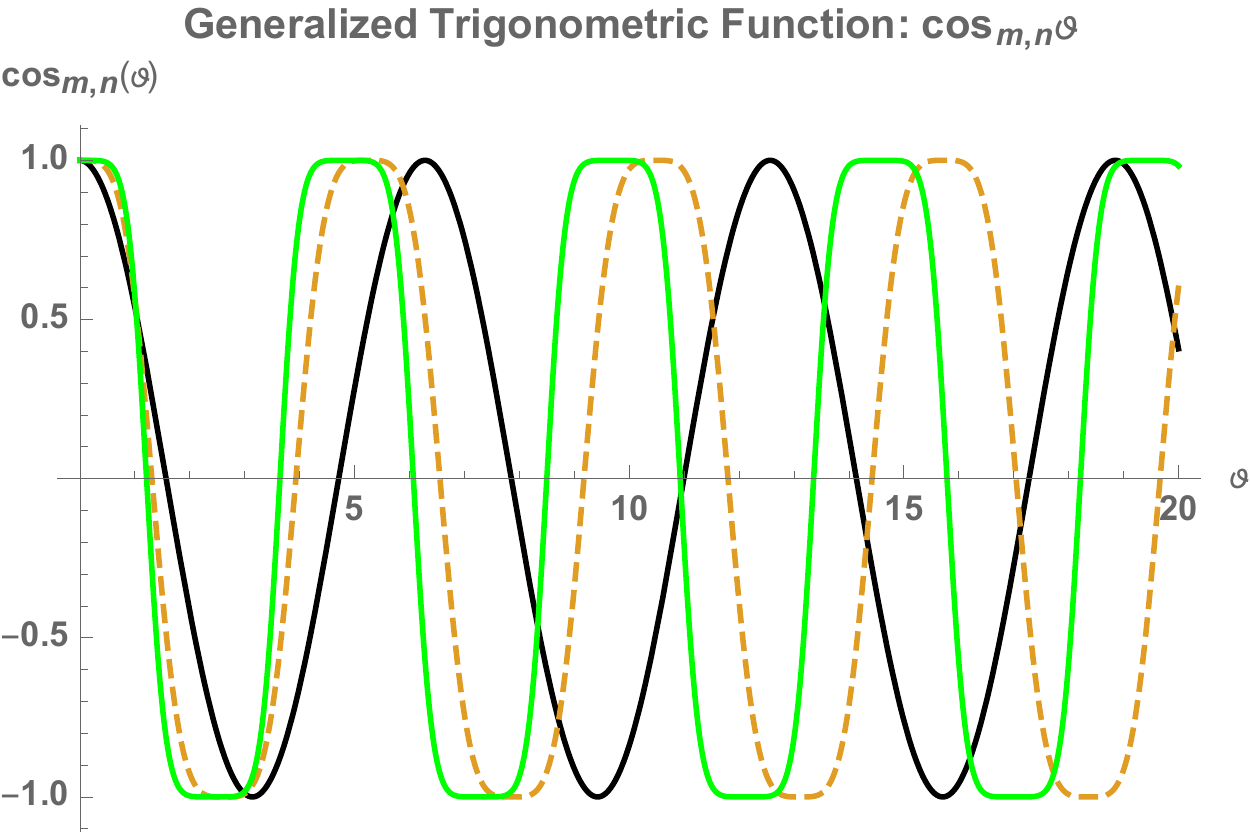}  		
		\caption{Plots of generalized trigonometric functions \(\cos_{2,\alpha+1}\) for $\alpha=1$ (Black), $\alpha=3$(Dashed), $\alpha=5$(Green)}   \label{cospq}		
	\end{center}
%	\end{wrapfigure}
\end{figure}

\begin{figure}[h]
%	 \begin{wrapfigure}{l}{0.3\textwidth}
	\begin{center}
		\centering
		\includegraphics[width=4.2in]{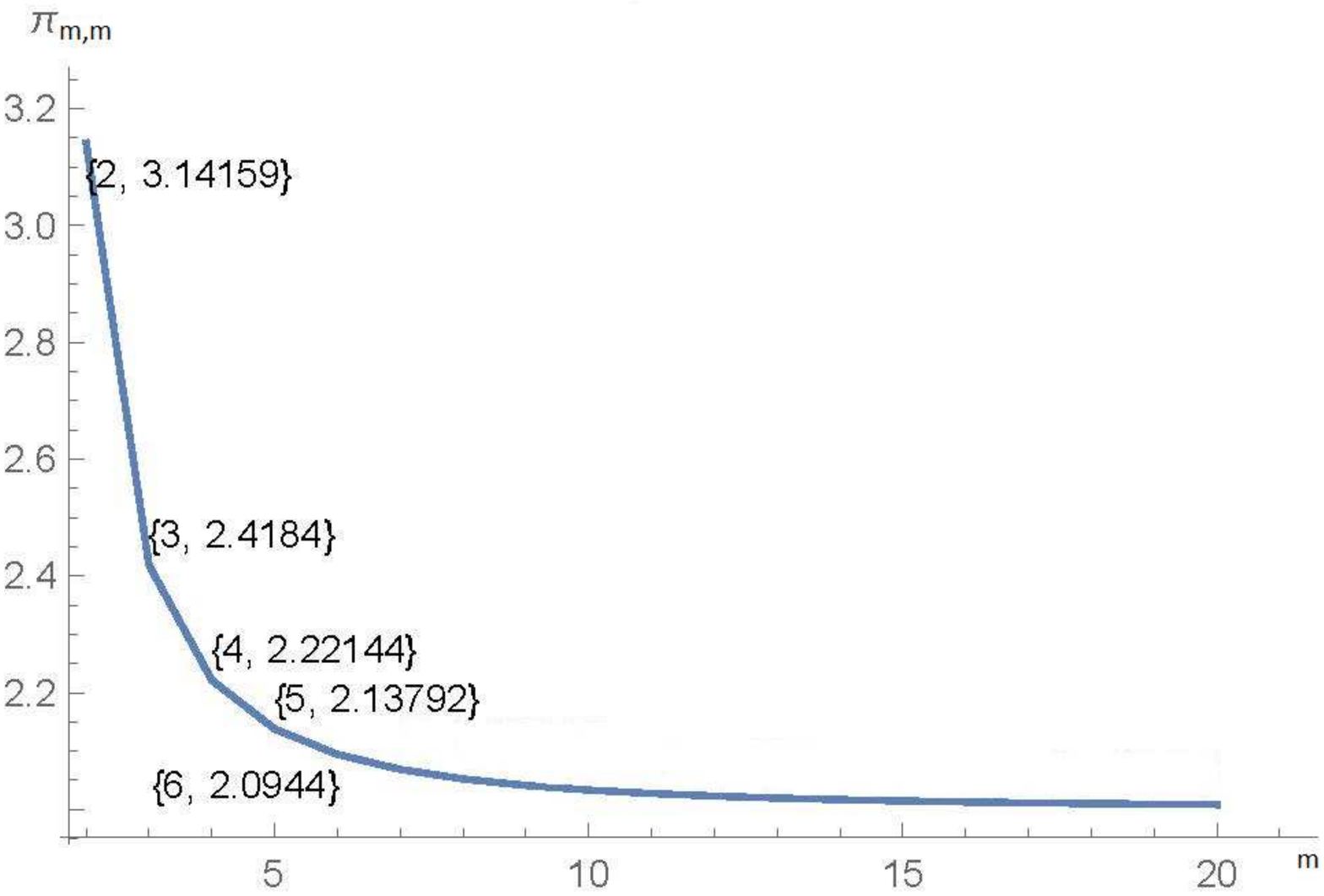}  		
		\caption{Plot of generalized $\pi_{m,n}$ for the special case $m=n$}   \label{pimm}		
	\end{center}
%	\end{wrapfigure}
\end{figure}

\begin{figure}[h]
%	 \begin{wrapfigure}{l}{0.3\textwidth}
	\begin{center}
		\centering
		\includegraphics[width=4.2in]{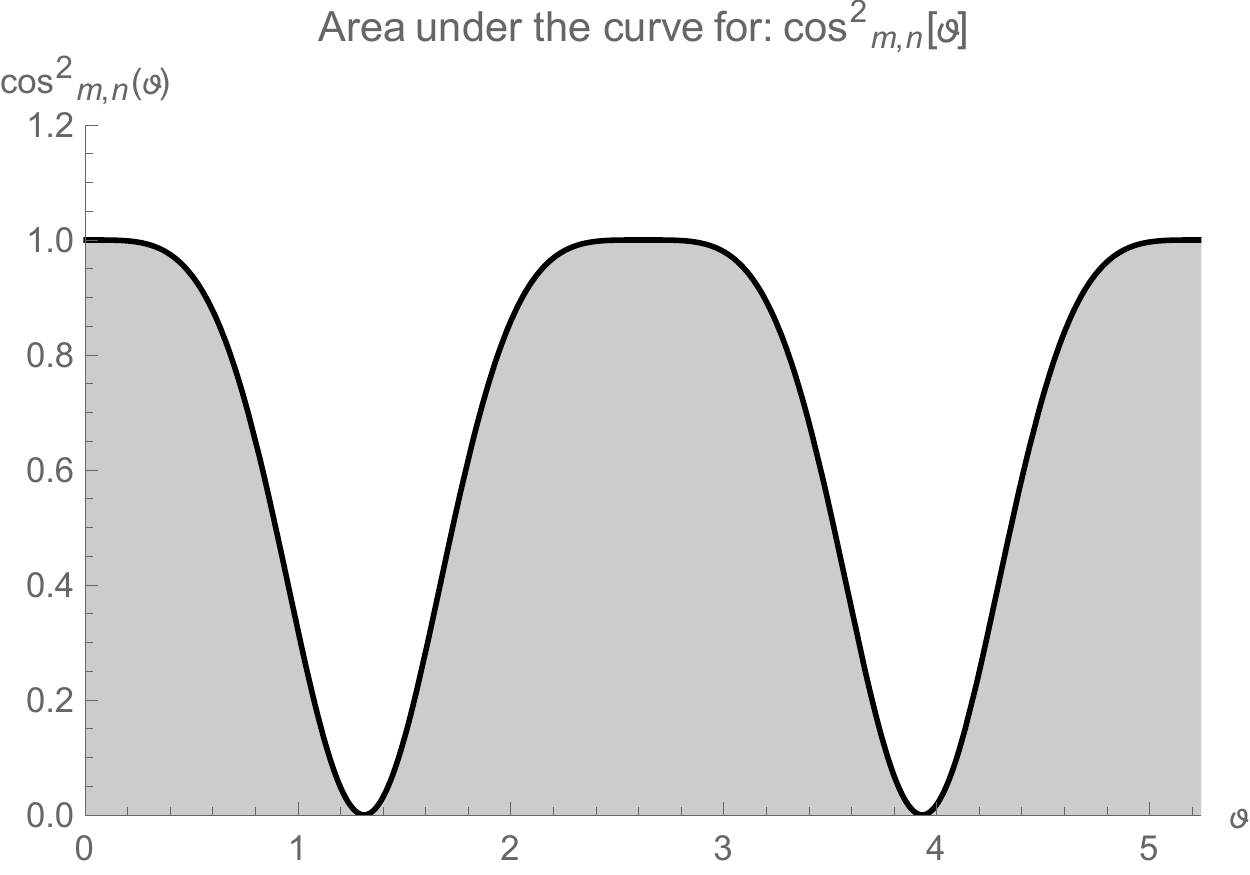}  		
		\caption{Sample plot of area under the curve for $\cos^2_{2,4}\theta$}   \label{area}		
	\end{center}
%	\end{wrapfigure}
\end{figure}

\begin{figure}[h]
%	 \begin{wrapfigure}{l}{0.3\textwidth}
	\begin{center}
		\centering
		\includegraphics[width=4.2in]{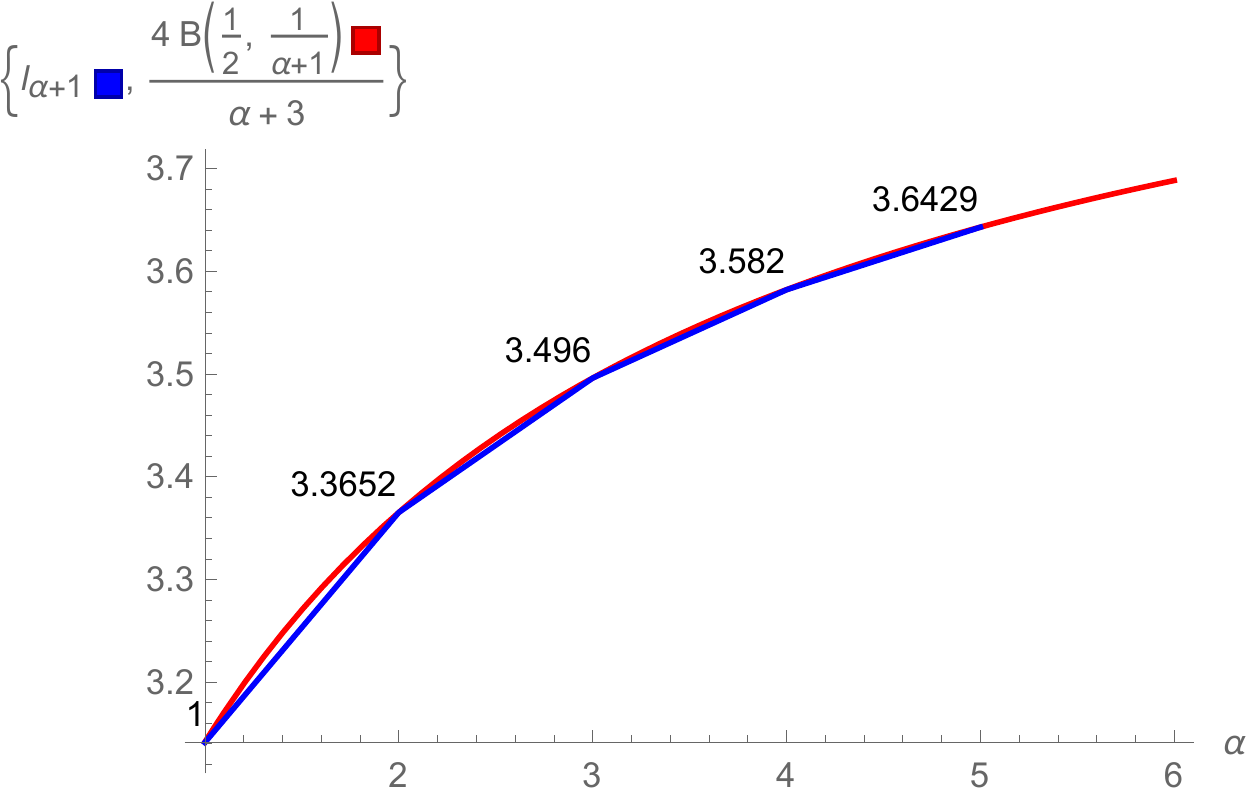}  		
		\caption{Comparative plot of L.H.S. (blue) and R.H.S. (red) of equation (\ref{2.1.5}) }   \label{compareIB}		
	\end{center}
%	\end{wrapfigure}
\end{figure}
Below, we give numerical proof that eqn. (\ref{2.1.5}) is correct. The R.H.S. of  eqn. (\ref{2.1.5}) is well known from the value of Beta function.  As for L.H.S, few generalized trigonometric functions can be plotted by numerical integration of the corresponding differential equations defining the functions. For details, we refer the reader to ref. [17] and directly present the plots in Figure-1, and a few of the generalized $\pi_{m,n}$ in Figure-2. The values of generalized $\pi_{m,n}$ are used to perform the integral in eqn. (\ref{2.1.4}) numerically. For instance,  Figure-3 shows the plot of area under the curve for $\cos^2_{2,\alpha+1}$ required to find the value $I_{\alpha+1}$ using numerical integral. For example, some of the values $I_{2}=3.14121,I_{3}=3.3652,I_{4}=3.496,I_{5}=3.582,I_{6}=3.6429$ agree (up to 3 decimal places) with the values on the right hand side of  eqn. (\ref{2.1.5}). This is seen clearly in the plot in Figure-(\ref{compareIB}). With this result and some manipulations one can write the action as,
\begin{equation}\label{2.1.6}
J =  \sqrt{\frac{32}{\alpha+1}} \frac{|q_0|^{\frac{\alpha+3}{2}}}{\alpha+3}B\bigg(\frac{1}{2},\frac{1}{\alpha+1}\bigg)
\end{equation}

\subsection{Period function and frequency}
The period function of the system can be easily obtained from the action:

\begin{equation}
  T(q_0) := \frac{\partial J}{\partial E}
\end{equation}

Simple manipulations give,

\begin{equation}
T(q_0)=\sqrt{\frac{8\pi}{\alpha+1}}
\frac{\Gamma\left(\frac{1}{\alpha+1}\right)}{\Gamma\left(\frac{\alpha+3}{2(\alpha+1)}\right)}|q_0|^{(1-\alpha)/2}
\end{equation}
which matches with the previously known result [6,7].

In expressing the solutions in form of generalized trigonometric functions, we recall the following relationship from [15]:

$$ \pi_{m,n} = \frac{2}{n}B\bigg(\frac{m-1}{m},\frac{1}{n}\bigg) $$

This means that one can express \(\pi_{2,\alpha+1}\) as,

$$ \pi_{2,\alpha+1} = \frac{2}{\alpha+1}B\bigg(\frac{1}{2},\frac{1}{\alpha+1}\bigg) $$

Now one can obtain an expression for the frequency of solution as,

$$\Omega_{2,\alpha+1} = \frac{2\pi_{2,\alpha+1}}{T(q_0)}$$

Simple manipulations give,

\begin{equation}
  \Omega_{2,\alpha+1} = \sqrt{\frac{2}{\alpha+1}}|q_0|^\frac{\alpha-1}{2}
\end{equation}

\subsection{Exact solutions}
Since we have defined the action angle variables in the previous section, the solutions of the Hamilton's equations can be obtained trivially. In this formalism the angle drops out of the hamiltonian and hence Hamilton's equations give, $$\dot{J} = 0$$ $$\dot{\theta} = \Omega_{2,\alpha+1}(J)$$

The first equation implies that \(J\) is a first integral. The second equation gives,

\begin{equation}
  \theta(t) = \Omega_{2,\alpha+1}t+\theta_0
\end{equation}

Hence, the solution of the purely nonlinear oscillator becomes,

\begin{equation}
  q(t) = q_0 \sin_{2,\alpha+1} (\Omega_{2,\alpha+1}t+\pi_{2,\alpha + 1}/2)
\end{equation}

Where from the initial condition \(q(0)=q_0\), we get $\theta_0 = \pi_{2,\alpha+1}/2$ since, $\sin_{m,n}(\pi_{m,n}/2) = 1$.

%%%%%%%%%%%%%%%%%%
\section{Remarks}
%%%%%%%%%%%%%%%%%%%
We conclude in this section with two important remarks and possible extensions of the current work.
First, in this work, we have obtained exact analytic solutions to the purely nonlinear oscillator. The solutions we presented were in terms of generalized trigonometric functions. As discussed in the Introduction, the purely nonlinear oscillator system we considered, admits solutions which have been written earlier in terms of ateb functions too. No known connection exists between the ateb and generalized trigonometric functions in literature as yet, as they are independent functions having different properties.  Since, we observed here that both the functions appear as solutions of the same nonlinear oscillator system, it is then natural to expect a relation between ateb and generalized trigonometric functions.   Although, we leave for later, a detailed investigation of the complete relation between the two functions, below we make two comments in support of their possible connection.  First, let us consider generalized trigonometric functons taking the values \(m=2\) and \(n=\alpha+1\);  their periodicity in this case takes the form:
 \begin{equation} \label{GTFpi}
 \pi_{2,\alpha+1} = \frac{2}{\alpha+1}B\bigg(\frac{1}{2},\frac{1}{\alpha+1}\bigg)\, .
 \end{equation}
Asymptotic approximations of ateb functions in terms of elementary functions have been considered in [19,20]. The periodicity of ateb functions is known to be:
 \begin{equation} \label{atebpi}
 \Pi_\alpha:=B\bigg(\frac{1}{\alpha+1}, \frac{1}{2}\bigg)\, .
 \end{equation}
From [eqn(\ref{GTFpi})] and [eqn(\ref{atebpi})], one may write:
\begin{equation}
 \pi_{2,\alpha+1} = \frac{2\Pi_\alpha}{\alpha+1}
\end{equation}
where one uses the property \(B(a,b)=B(b,a)\). Thus, the $2\Pi_\alpha$ periodicity of ateb functions and $2\pi_{2,\alpha+1}$ periodicity of generalized trigonometric functions with \(m=2; n= \alpha+1\) are same, up to a factor of $2/(1+\alpha)$. Second, we recall from [18] the following relation involving the generalized trigonometric functions:
\begin{equation}
  \arcsin_{m,n} x := \frac{1}{n}B_{x^n} \bigg(\frac{1}{n},\frac{m-1}{m}\bigg), \hspace{10mm} x \in [0,1]\, ,
\end{equation}
where $B_{x}(a,b)$ is the incomplete beta function.  The inversion of incomplete beta function is actually the ateb function, there by giving a way to obtain generalized trigonometric functions. These connections may be investigated further to learn whether the generalized trigonometric functions and the periodical ateb functions also share similar identities, and in particular, whether a deeper mathematical connection exists. These works are in progress.\\

\noindent
Second, we make some remarks on the importance of pursuing action-angle variables for purely nonlinear oscillators and open issues. One reason is the possibility of extending the results to situations where the system becomes time-dependent and quasi-periodic. For instance, if the length of the pendulum changes ever so slowly, it is important to know which physical quantities of the system remain invariant. It is well known that for standard linear harmonic oscillator, the action variable of the Hamiltonian of the system remains unchanged during a period of oscillation and is an adiabatic invariant. There are in fact, exact invariants, such as the Ermakov-Lewis invariant, which remain unchanged even under more general time-dependent situations~\cite{Goodball}. A proper proof of the adiabatic invariance of the classical action was constructed in~\cite{wells} for a general Harmonic oscillator. Our aim is to check the whether the action variable $J$ proposed here is such an invariant for the corresponding nonlinear oscillator. Time-dependent Hamiltonian systems are ubiquitous in science and engineering and hence the question is well motivated. Following the general idea of action-angle variables,  a new pair of coordinates $(I,\phi)$ given as:
\begin{equation}
I =  \left(\frac{p^2}{2\tau} + \frac{|q|^{\alpha + 1}}{(\alpha + 1)\tau} \right)\equiv H\tau, \quad
\phi =\arcsin_{2,\alpha+1}  \frac q {\sqrt{p^2 \tau^2 + 2\,\frac{|q|^{\alpha + 1}}{\alpha + 1}}} \equiv \arcsin_{2,\alpha+1}\frac q
     {\sqrt{2H\tau^2}} \, ,
\label{Iphi}
\end{equation}
may be proposed. Here $H$ is given in eqn. (\ref{HNA}) and $I$ is a generalized momentum related to the putative adiabatic invariant given by $J$, defined earlier as
\begin{equation}
  J = \oint I \, d\phi \, .
\end{equation}
If $q$ and $p$ evolutions can be understood from the Hamilton's equations following from $H$, then $I$ and $\phi$ are expected to satisfy the new Hamilton's equations following from the new Hamiltonian $K$, which is generally obtained via a canonical transformation from $H$ as:
\begin{equation} \label{Kamiltonian}
K-H =\left( \frac{\partial F(I,q,\tau)}{\partial \tau}\right)_{I,q} =  \dot\tau \left( \frac{\partial F(I,q,\tau)}{\partial t}\right)_{I,q}  \,
\end{equation}
Here, $F$ is the generating function given by:
\begin{equation}\label{gen}
F(I,q,\tau)=\int_0^q p \,dq',
\end{equation}
With the set up as above, the proof of adiabatic invariance of the action given in~\cite{wells}, depends on the crucial condition that the generating function $F(I,q,\tau)$ is independent of $\dot\tau(t)$ (see the discussion after eqn. 34 in ~\cite{wells}). If this condition is satisfied, then it was shown in~\cite{wells} that $|J-I| = O(\dot\tau)$ and also $dJ/dt= O(\ddot\tau, \dot\tau^2)$, signifying that the action variable $J$ varies more slowly than the slowly varying parameter $\tau(t)$. It is in this precise sense that the $J$ is an adiabatic invariant. The generating function alluded to in eqn. (\ref{gen}) can be computed for purely nonlinear oscillator, to be:
\begin{eqnarray}
&& F(I,q,\tau) =\int_0^q\sqrt{2I/\tau-q^{\alpha +1}/\tau^2}\,d q \nonumber\\
&& =-q \frac{\tau \left((\alpha +1) \text{I} \sqrt{4-\frac{2 q^{\alpha +1}}{\text{I}\, \tau}} \, _2F_1\left(\frac{1}{2},\frac{1}{\alpha +1};1+\frac{1}{\alpha
   +1};\frac{q^{\alpha +1}}{2 \text{I}\, \tau}\right)+4 \text{I}\right)-2 q^{\alpha +1}}{(\alpha +3) \tau^2 \sqrt{\frac{q^{\alpha +1}-2 \text{I} \,\tau}{\tau^2}}} \, .
\end{eqnarray}
It can now be shown that the new Hamiltonian for purely nonlinear oscillator written in terms of the action $I$ is $K =  I/\tau + \dot\tau \left( \frac{\partial F}{\partial t}\right)_{I,q} $, where
\begin{equation} \label{kam}
\left( \frac{\partial F}{\partial t}\right)_{I,q}=  - \frac{q \left(\tau \left((\alpha -1) \text{I} \sqrt{4-\frac{2 q^{\alpha +1}}{\text{I} \tau}} \, _2F_1\left(\frac{1}{2},\frac{1}{\alpha
   +1};1+\frac{1}{\alpha +1};\frac{q^{\alpha +1}}{2 \text{I} \tau}\right)+8 \text{I}\right)-4 q^{\alpha +1}\right)}{2 (\alpha +3) \tau^3 \sqrt{\frac{q^{\alpha
   +1}-2 \text{I} \tau}{\tau^2}}}
\end{equation}
is clearly independent of $\dot\tau$. As this key condition is satisfied, the proof given in~\cite{wells} goes through, with the result that the action variable $J$ for purely nonlinear oscillator is an adiabatic invariant. As a consistency check, for $\alpha=1$, it can be shown that the new generating function in eqn (\ref{kam}), reduces to the one given in~\cite{wells}, where for the case of standard linear oscillator, the adiabatic invariance of the action has been verified.  From science and engineering applications point of view, the full conversion of the new Hamiltonian $K$ in eqn (\ref{Kamiltonian}) and generating function $F$ in  eqn (\ref{kam}), in terms of the generalized trigonometric functions (using eqn 3.2) needs to be done, possibly using the methods in~\cite{takeuchi}. We leave these issues for future work.

\section*{Acknowledgements}
The authors would like to thank the anonymous referees for their valuable comments that led to the improvement of the article.

\noindent

\appendix
\section*{Appendices}
\addcontentsline{toc}{section}{Appendices}
\renewcommand{\thesubsection}{\Alph{subsection}}

\subsection{Action-angle variables}
\renewcommand{\theequation}{\thesubsection.\arabic{equation}}
Consider a one-dimensional Hamiltonian system that is independent of time,
\bea \label{Ham}
H(q,p) = \frac{p^2}{2m} + V(q)\, ,
\eea
with generalized coordinates $(q,p)$, whose evolution is given by the level curves of $H$.
\begin{figure}[h]
%	 \begin{wrapfigure}{l}{0.3\textwidth}
	\begin{center}
		\centering
		\includegraphics[width=2.2in]{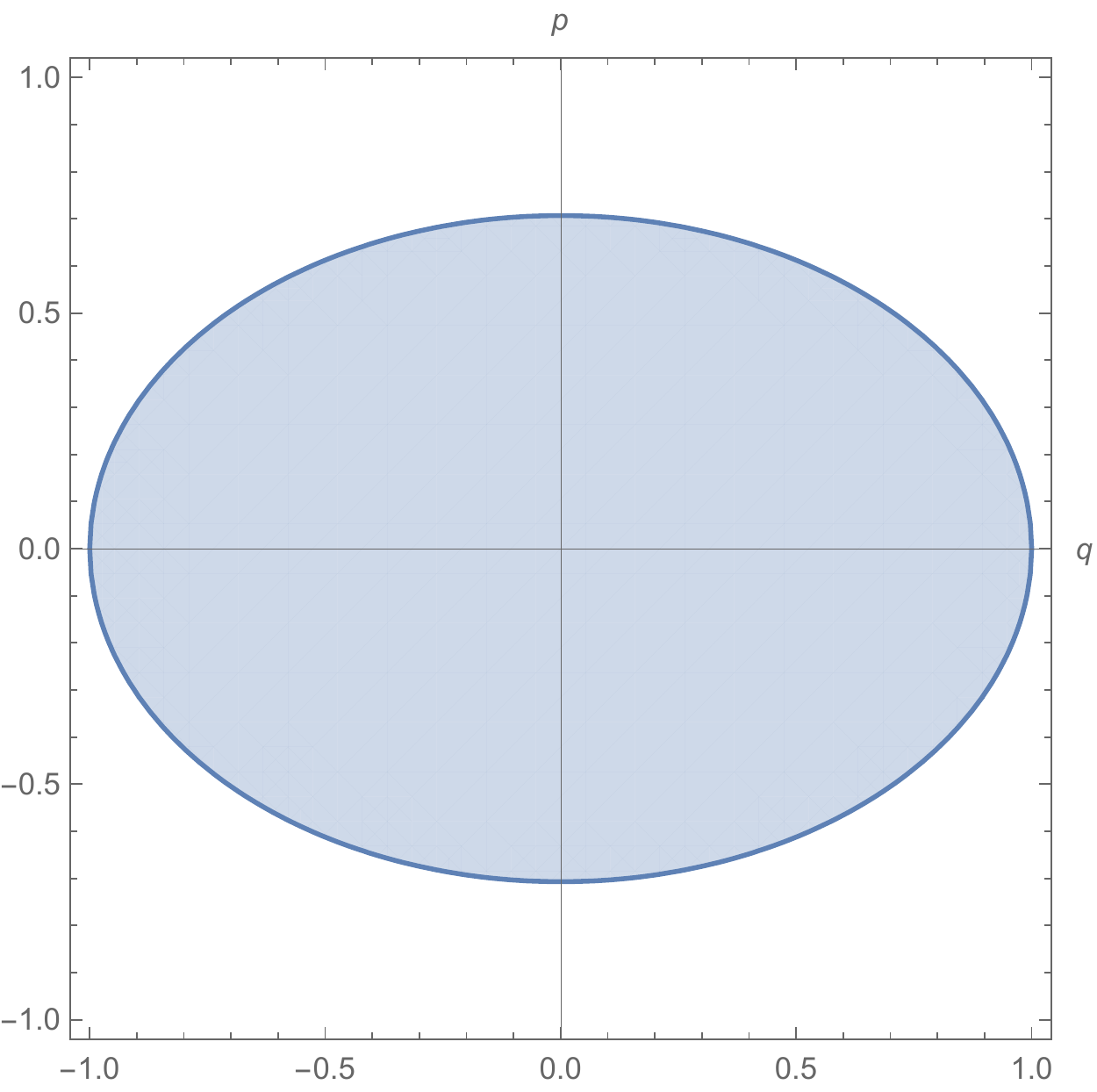}  		
		\caption{A phase portrait corresponding to a simple harmonic oscillator}   \label{ellipse}		
	\end{center}
%	\end{wrapfigure}
\end{figure}
A typical phase portrait is given in Figure-\ref{ellipse}, where the motion corresponds to traversing along the boundary of the curve (ellipse in this case).  The families of curves and the path in the phase portrait depends on a chosen value of the constant of motion. With the assumption that the potential energy $V(q)$ allows the system to perform periodic motion between turning points $q_{\pm}$, one can naturally get an angle by parameterization of the curve. If the constant of motion is denoted by G and its value by $g$ with the path given by $\gamma(g)$, the action then becomes the line integral enclosing the area inside the curve:
\bea \label{action}
I(g) = \frac{1}{2\pi}\oint_{\gamma(g)}\, p\, dq \, .
\eea
It is useful have an example in mind, such as the harmonic oscillator having the Hamiltonian $H(q,p)= p^2/{2m} + kq^2/2$. We proceed in two methods.\\

\noindent
{\underline{Method-I}}: For the Harmonic oscillator, the solutions for generalized coordinates known to be:
\bea \label{sol}
q(t) = A \sin\phi(t), \quad p(t) = m\omega A \cos\phi(t),
\eea
where the phase $\phi(t) = \omega t + \phi_0$ giving us a natural angle variable, which keeps track of the position on the ellipse in figure-(\ref{ellipse}). Here, $\omega = \sqrt{k/m}$, $A$ stands for the amplitude and $\phi_0$ for the initial phase. The action variable can now be found using the integral in eqn.(\ref{action}) to be
\bea
I = \frac{1}{2\pi}\oint_{\gamma}\, p\, dq \, = \frac{1}{2\pi}\int\limits_0^{2\pi/\omega} \, m\omega A \cos\phi(t) d(A\sin\phi(t)) = \frac{1}{2} m \omega A^2.
\eea
This suggests that a transformation to new variables is possible as: $(q,p) \rightarrow (\phi, I)$, with angle being the phase of the oscillator and action standing for the amplitude. The Hamiltonian can be rewritten in terms of these new action-angle variables, from the fact that Energy (constant) at a turning point is $E = \frac{1}{2}k A^2 = \omega I = H(I)$. The motion now becomes simple, as the angle $\phi$ drops out of the Hamiltonian.  \\

\noindent
{\underline{Method-II}}:
The reason for employing action-angle variables in this manuscript is for an entirely different reason, which is, that an explicit knowledge of the solutions in eqn.(\ref{sol}) is generally  not required (irrespective of whether analytical solutions can be found out). This is so because (for conservative systems where $H=E$ is fixed), following eqn. (\ref{Ham}), the generalized momentum can always be written as
\bea
p = \sqrt{2 m \left(E - V(q)\right)} \, .
\eea
Even without explicit knowledge of solutions for $(q,p)$, the action variable can always be defined as
\bea \label{m2}
I = \frac{1}{2\pi}\oint_{\gamma}\, p\, dq \, = \frac{2 m}{\pi}\int\limits_{-\sqrt{2E/k}}^{\sqrt{2E/k}} \,  \sqrt{2 m \left(E - \frac{1}{2} k q^2\right)} dq \, = \frac{E}{\omega} .
\eea
The result in eqn. (\ref{m2}) matches the one found using Method-I above. The only information used in Method-II is the shape of the level curve. Furthermore, as mentioned in the introduction, an important reason for the Action-Angle method is its application to situations where the Hamiltonian of the dynamical system is time-dependent. In this situation, the action-angle method gives rise to adiabatic invariants [23], which do not change and are important in checking the integrability of the model in question.

As in the harmonic oscillator case, one can express the frequency of a general system as, $$\omega = \frac{\partial E}{\partial I}$$

Defining \(J = 2\pi I\) it is easy to see that the time period of oscillation,

\begin{equation}
  T = \frac{\partial J}{\partial E}
\end{equation}

This is form of the action that we have used in section 2 and holds for the general case even where the motion is not necessarily harmonic. In the harmonic oscillator case discussed above we showed that transforming to the action-angle variables makes the Hamiltonian independent of the angular coordinate. This holds true even for higher degrees of freedom making the Hamilton's equations trivial: the actions are constants of motion while the angles evolve linearly in time. The action-angle formalism [21,22] can be readily extended to systems with higher degrees of freedom where sufficient constants of motion exist and the Hamilton-Jacobi equation is separable. A general theorem in classical dynamics guarantees the existence action-angle variables for an \(n\) dimensional system if there exist at least \(n\) independent constants of motion in involution (see for example [22] for a more detailed account).

\subsection{Generalized Trigonometric Functions}
\renewcommand{\theequation}{\thesubsection.\arabic{equation}}

Consider the general curve of the following form:
\begin{equation}
  |x|^m + |y|^n = 1, \hspace{10mm} m,n \in (1,\infty)
\end{equation}

Figure-6 shows plots of three such curves for three different pairs of \(m\) and \(n\). For the case \(m=n=2\) we get a standard circle and by analogy to the standard circle, one can define parametric equations to such a "generalized" circle to be the so called "generalized trigonometric functions". They are $y = \sin_{m,n} \theta$ and $x = \cos_{m,n} \theta$. We remark that the parameters \(m\) and \(n\) are completely independent in the most general case. They are rational numbers and can take any value greater than one. Figure-7 shows plots of \(\sin_{m,n} \theta\) for two different pairs of \(m\) and \(n\) values.

\begin{figure}[h]
%	 \begin{wrapfigure}{l}{0.3\textwidth}
	\begin{center}
		\centering
		\includegraphics[width=3.2in]{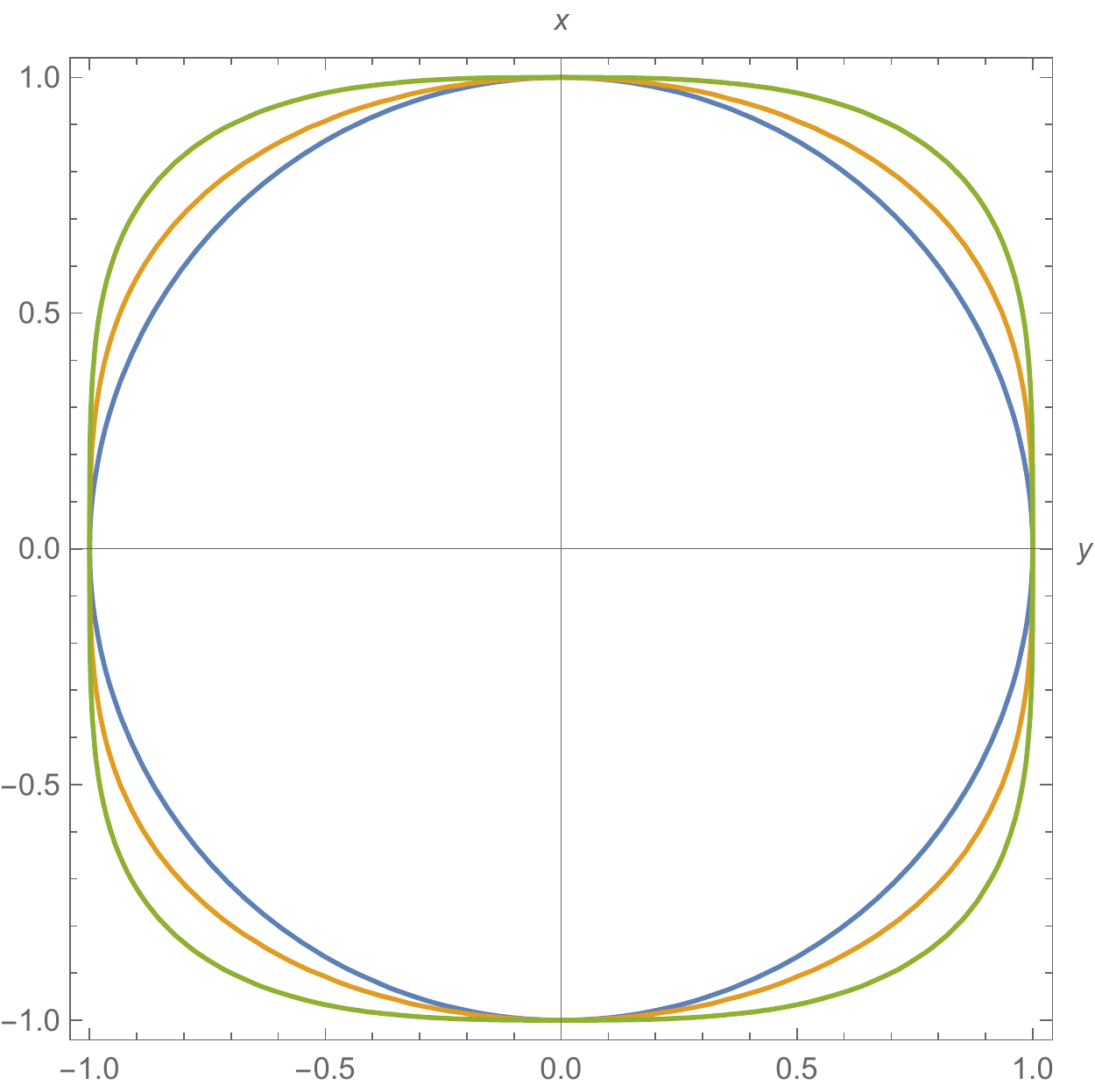}  		
		\caption{Plot of $|x|^m + |y|^n = 1$ with  \(m=2, n=2\) (Blue, circle), \(m=2, n=3\) (Orange) and \(m=3, n=4\) (Green) }   \label{generalized curves}		
	\end{center}
%	\end{wrapfigure}
\end{figure}

\smallskip

Clearly, they satisfy the identity
\begin{equation}
  |\cos_{m,n} \theta|^m + |\sin_{m,n} \theta|^n = 1
\end{equation}

These functions exhibit periodic behaviour similar to the standard trigonometric functions. A rigourous definition for the generalized trigonometric functions may be given as follows. We define the \(\sin_{m,n} x\) as the inversion of the integral, \begin{equation} x = \int_{0}^{y} \frac{dy'}{\sqrt[m]{1-|y'|^n}}\end{equation} This integral \(x=x(y)\) is monotonically increasing and hence can be inverted so that we get \begin{equation}y(x) := \sin_{m,n} x, \hspace{10mm} x \in [0,\pi_{m,n}/2]\end{equation} Defined initially in \([0,\pi_{m,n}/2]\), the domain can be extended throughout the real line by a natural continuation process. These functions are \(2\pi_{m,n}\) periodic with, \begin{equation}\pi_{m,n} := 2\int_{0}^{1} \frac{dx'}{\sqrt[m]{1 - |x'|^n}}\end{equation}

\begin{figure}[h]
%	 \begin{wrapfigure}{l}{0.3\textwidth}
	\begin{center}
		\centering
		\includegraphics[width=4.2in]{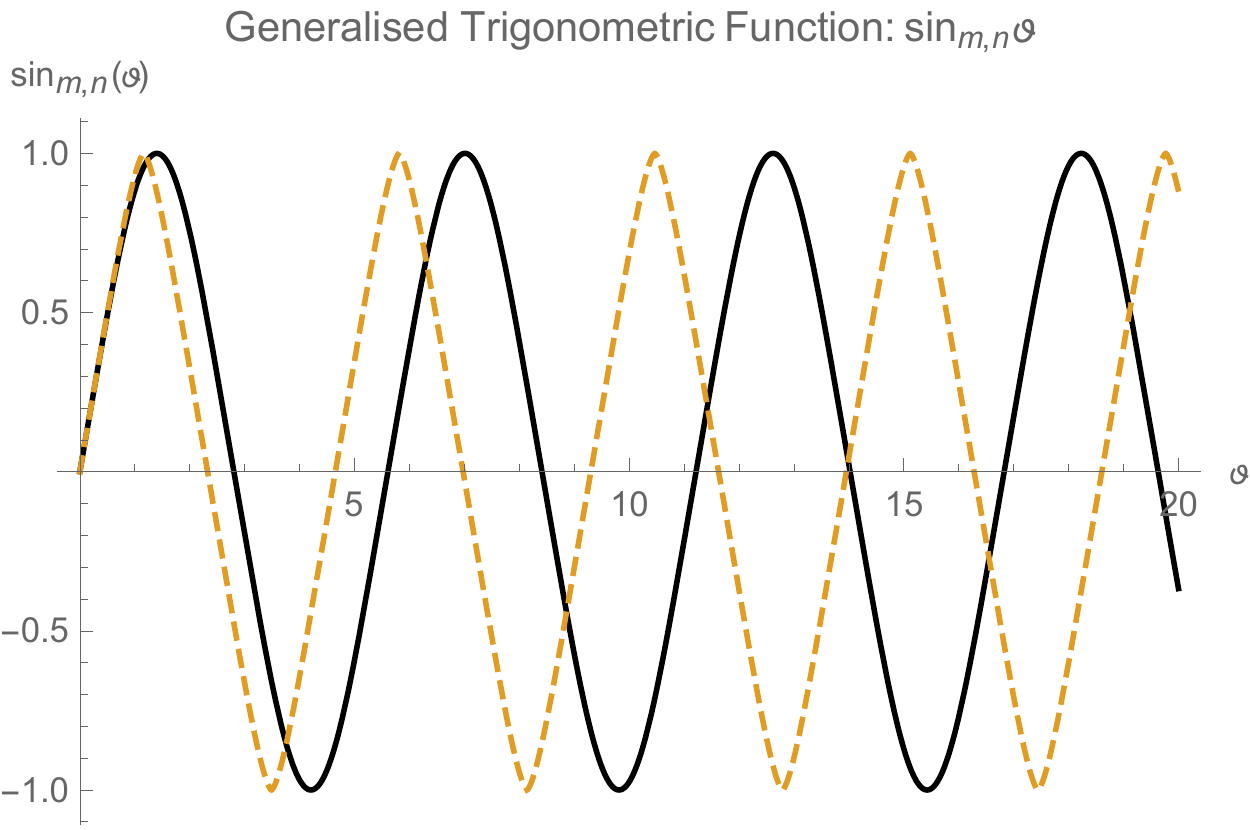}  		
		\caption{Plot of $\sin_{m,n} \theta$ with \(m=2, n=3\) (Black) and \(m=3, n=4\) (Dashed) }   \label{sine}		
	\end{center}
%	\end{wrapfigure}
\end{figure}

We define the corresponding cosine function as, \begin{equation}
                                                  \cos_{m,n} x := \frac{d}{dx}(\sin_{m,n} x)
                                                \end{equation}

The following basic properties hold:$$\sin_{m,n} (0) = 0$$ $$\cos_{m,n} (0) = 1$$ $$\sin_{m,n} (\pi_{m,n}/2) = 1$$  $$\cos_{m,n} (\pi_{m,n}/2) = 0$$

These functions clearly show a resemblance with the familiar trigonometric functions. It is very easily verified that for \(m=n=2\), all of these definitions and results reduce to those of ordinary trigonometric functions. It should however be remarked that it is not necessary that all the properties of the standard trigonometric functions be carried out to their generalized counterparts. See refs [14-18] for detailed discussions on these functions.

\end{document}